\documentclass[graybox]{svmult}
\usepackage[
backend=biber,
style=numeric,
]{biblatex}
\usepackage{lmodern}
\usepackage{graphicx}
\usepackage{amsmath}
\usepackage{quotes}
\usepackage{algorithm}
\usepackage{amssymb}
\usepackage[noend]{algpseudocode}
\usepackage{changepage}
\usepackage{rotating}
\usepackage{subcaption}

\usepackage{hyperref}
\usepackage{xcolor}
\definecolor{goodblue}{RGB}{0, 91, 187}
\hypersetup{
  colorlinks=true,
  allcolors=goodblue,
  urlcolor=goodblue,
  citecolor=goodblue,
  pdfborder={0 0 0},
  breaklinks=true,
}
\usepackage{booktabs}

\usepackage{bookmark}
\makeatletter
\providecommand*{\toclevel@titlech}{0} 
\edef\toclevel@authorch{\the\numexpr\toclevel@titlech+1} 
\makeatother

\newcommand{\sr}[1]{\sigma_{\text{SR}}(#1)}
\newcommand{\simppair}[1]{\text{sp}(#1)}
\newcommand{\msr}[1]{M_{\text{SR}}(#1)}
\newcommand{\hypergraph}{\mathcal{H}}

\newcommand{\bS}{\mathbf{S}}
\newcommand{\ssf}{\sigma_{\mathrm{SF}}}
\newcommand{\ses}{\sigma_{\mathrm{ES}}}
\newcommand{\sfes}{\sigma_{\mathrm{FES}}}
\newcommand{\des}{d_{\mathrm{ES}}}
\newcommand{\dnes}{d_{\mathrm{NES}}}

\newcommand{\Etilde}{\widetilde{E}}
\newcommand{\HSC}{\mathcal{H}^{\text{SC}}}
\newcommand{\HCM}{\mathcal{H}^{\text{CM}}}
\newcommand{\graph}{\mathcal{G}}
\newcommand{\GSC}{\mathcal{G}^{\text{SC}}}
\newcommand{\GCM}{\mathcal{G}^{\text{CM}}}
\newcommand{\ESC}{E^{\text{SC}}}
\newcommand{\ECM}{E^{\text{CM}}}
\newcommand{\DSC}{D^{\text{SC}}}
\newcommand{\DCM}{D^{\text{CM}}}
\newcommand{\din}{d^{\text{in}}}
\newcommand{\dout}{d^{\text{out}}}

\newcommand{\ps}[1]{\mathcal{P}(#1)}
\newcommand{\srps}[2]{\mathcal{P}_{#2}(#1)}

\title{The nestedness of higher-order networks}
\titlerunning{The nestedness of higher-order networks}
\authorrunning{T. LaRock, Y. Zhang, J.-G. Young, N. Eikmeier, R. Lambiotte, and N. Landry}
\author{Timothy LaRock\orcidID{0000-0003-0801-3917}\\
Yanting Zhang\orcidID{0009-0001-8039-2408}\\
Jean-Gabriel Young\orcidID{0000-0002-4464-2692}\\
Nicole Eikmeier\orcidID{0000-0002-4789-2275}\\
Renaud Lambiotte\orcidID{0000-0002-0583-4595}\\
Nicholas W. Landry\orcidID{0000-0003-1270-4980}
}
\institute{Timothy LaRock \at Department of Civil and Environmental Engineering, Princeton University, \email{larock@princeton.edu}
\and
Yanting Zhang \at Institute of Data Science, University of Hong Kong \email{yantingz@connect.hku.hk}
\and
Jean-Gabriel Young \at Vermont Complex Systems Institute, University of Vermont, \email{jean-gabriel.young@uvm.edu} \at Department of Mathematics and Statistics, University of Vermont
\and
Nicole Eikmeier \at Department of Computer Science, Grinnell College, \email{eikmeier@grinnell.edu}
\and
Renaud Lambiotte \at Mathematical Institute, University of Oxford, \email{renaud.lambiotte@maths.ox.ac.uk} 
\and
Nicholas W. Landry \at Department of Biology, University of Virginia, \email{nicholas.landry@virginia.edu} \at School of Data Science, University of Virginia \at Vermont Complex Systems Institute, University of Vermont
}

\begin{document}

\maketitle

\abstract{
In contrast to dyadic interactions, higher-order interactions may contain one another, with subgroups naturally embedded within larger groups. These containment patterns arise empirically in ecology, sociology, computer science and the science of science, and have been studied under the names nestedness, simpliciality, encapsulation, and inclusion. In this chapter, we review each of these measures and unify them through a mathematical object known as the encapsulation directed acyclic graph, formulating each measure as a function of its properties. We demonstrate that nested structure is prevalent in social systems across several domains, show that different measures capture complementary aspects of this structure, and find that the absence of nestedness can itself be a powerful indicator of the mesoscale organization of a system.
}

\section{Introduction}

A natural consequence of modeling complex systems as collections of group interactions is that smaller interactions can be fully contained within larger ones. 
We will refer to this property as \emph{nestedness}.
Nestedness is a characteristic feature in many social systems.
For example, there is a rich history of measuring nestedness in ecological systems; species distribution patterns~\cite{crowell_comparison_1986} and mutualisms~\cite{bascompte_nested_2003} often exhibit nestedness, which is thought to be due to non-random processes such as selective extinction~\cite{patterson_nested_1986}.
In human social systems, social contacts as measured by proximity can be highly nested~\cite{landry_simpliciality_2024} because of the high density of these interactions and non-random fractures and aggregation of groups~\cite{iacopini_temporal_2024}.
In organizational communication, such as email or instant messaging, nested interactions may simultaneously reflect team structure and organizational hierarchy.
Email exchanges, for example, where a higher-order interaction comprises the sender and the recipients of an email, often exhibit nested structure~\cite{larock_encapsulation_2023} depending on the particular measure of nestedness~\cite{landry_simpliciality_2024}.

Nestedness in human social systems is both nuanced and contextual as it may manifest in counterintuitive or asymmetric ways.
For instance, while triads of co-authors who have each collaborated in pairs strongly predict a joint three-author publication \cite{patania_shape_2017}, the converse does not hold: co-authorship networks contain little nestedness overall, relative to the maximum possible \cite{larock_encapsulation_2023}.
This asymmetry may reflect the high cost of collaboration; unlike emails or fleeting encounters, co-authorships often span years and require sustained investment, making it unlikely that a strict subset of co-authors would publish independently from the larger whole.
More broadly, the temporal ordering of interactions appears to play a key role: small groups may merge into larger ones over time, but larger groups rarely fragment into active smaller subgroups. Recent work on the merging and splitting dynamics of groups reinforces this view, revealing rich temporal patterns tied to nestedness \cite{iacopini_temporal_2024}.
The presence\textemdash{}or absence\textemdash{}of nestedness thus offers direct insight into the generative dynamics of a social system.

Nestedness also affects dynamical processes on higher-order systems. Increased nestedness lowers the epidemic threshold, produces smaller outbreaks, and delays the onset of bistability in contagion models \cite{kim_contagion_2023,burgio_triadic_2024,malizia_nested_2026}. It also delays the onset of synchronization \cite{zhang_higherorder_2023}, and there exists an optimal level of nestedness that maximizes synchronizability for certain ratios of 2- and 3-hyperedges.

The structural phenomenon of nestedness has been referred to in various ways across scientific fields.
Early work in ecology introduced nestedness to quantify species distribution patterns by determining whether assemblage of species at specific cites were subsets of groups found at richer sites.~\cite{patterson_nested_1986,atmar_measure_1993}.
This measure, broadly speaking, quantifies how ``packed'' an incidence matrix is; a perfectly nested matrix is one in which every column is a proper subset of the preceding column. However, this measure is not guaranteed to be unique and can also be expensive to compute~\cite{guimaraes_improving_2006}. In addition, this measure does not naturally capture the complex structure of nested interactions from a higher-order network perspective, and we therefore exclude it from our overview of nestedness measures for the remainder of this chapter.

When one edge is a subset of another, we say the larger edge includes the smaller one, and refer to them as the containing and contained edges, respectively. At the system level, the collection of all such relationships defines the inclusion structure of a hypergraph.
In the network science community, Ref.~\cite{larock_encapsulation_2023} defined \textit{encapsulation} as a directed graph comprising the subset relationships contained in a hypergraph~\cite{larock_encapsulation_2023}.
Ref.~\cite{landry_simpliciality_2024} explicitly compared different properties of a hypergraph to a representative simplicial complex, naming the resulting family of measures \textit{simpliciality}.
In contrast, Ref.~\cite{barrett_counting_2025} defined the term \textit{simplicial pairs} as a way to quantify whether two selected edges contain each other more than one might expect at random.
Ref.~\cite{lotito_higherorder_2022} quantified nestedness through \textit{motifs}, defined as small connected subgraphs containing hyperedges of arbitrary sizes, and \textit{nested structure}, defined as the number of subsets of a single hyperedge.
We describe each of these measures in detail below.

While we focus on nestedness in this chapter, it is a special case of hyperedge overlap, where two hyperedges share at least one node in common.
Nestedness imposes the additional constraint that the two hyperedges share all of the nodes of one of the hyperedges; relaxing this constraint can also yield rich insights into the higher-order organization of a system.
Refs.~\cite{aksoy_hypernetwork_2020} and \cite{chodrow_configuration_2020} define the \textit{higher-order $s$-walk} and \textit{intersection profile} for hypergraphs as tools to quantify the distribution of overlapping interactions in a system.

The remainder of this chapter is organized as follows. We first introduce the encapsulation directed acyclic graph (DAG), its structural properties, and its relationship to other mathematical representations.
Second, we review existing measures of nestedness in hypergraphs and unify them by expressing them as structural properties of the encapsulation DAG.
For each measure, we discuss its local and global variants, limitations, and computational considerations.
Third, we review existing generative higher-order network models and quantify the nestedness how well they can tune nestedness.
Lastly, we present a case study quantifying the nestedness of various social systems across several domains and discuss the implications of different measures of nestedness on the structure of these systems.

\section{The encapsulation DAG}

We represent the nested structure of higher-order interactions using \emph{line graphs} \cite{evans_line_2009}, which map hyperedges to nodes and relationships between hyperedges to edges~\cite{aksoy_hypernetwork_2020,lambiotte_networks_2019}. Here, we focus on a specific type of line graph that captures subset relationships: each directed edge points from a containing hyperedge to the hyperedge it contains. The resulting structure is necessarily acyclic and is known as the \emph{encapsulation directed acyclic graph (DAG)}, introduced in Ref.~\cite{larock_encapsulation_2023}. It is defined as follows:

\begin{definition}{\textit{Encapsulation DAG} (\cite{larock_encapsulation_2023}).}
Given a simple hypergraph $\hypergraph = (V, E)$, $\graph = (E, D)$ is the hypergraph's encapsulation DAG, where each node in $\graph$ corresponds to a hyperedge in $E$ and $D$ denotes the set of directed edges, with $(e_i, e_j) \in D$ if $e_i, e_j \in E$, and the nodes of $e_j$ are fully contained within the nodes of $e_i$, i.e., $e_j \subset e_i$.
\end{definition}
Since $|e_i| \neq |e_j|$ for every connected $e_i$ and $e_j$ in an encapsulation DAG, a cycle in this graph would imply that a smaller hyperedge includes a larger hyperedge, which is impossible, and so the graph is always a DAG \cite{larock_encapsulation_2023}.

\begin{figure}[ht!]
    \centering
    \includegraphics[width=\linewidth]{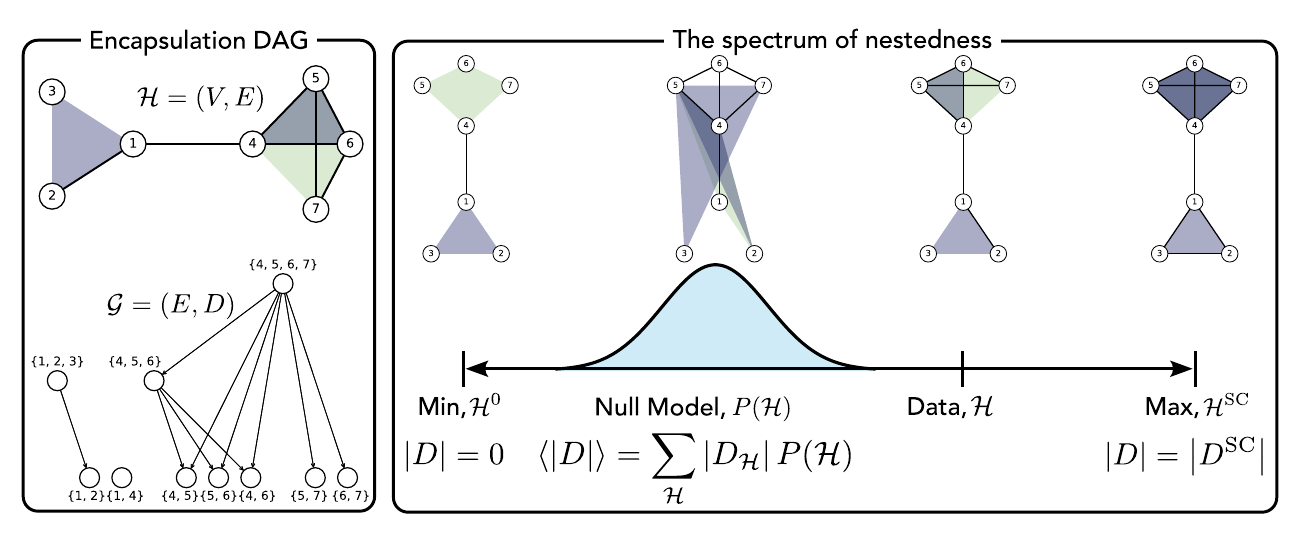}
    \caption{\textbf{An illustration of our unified nestedness framework based on the encapsulation DAG.}
    The left panel illustrates a hypergraph with its associated encapsulation DAG, which captures the hierarchical subset relations between hyperedges. In the latter, the number of nodes (hyperedges) is $|E|$ and the number of directed relations is $|D|$.  
    Hyperedges are arranged vertically by their size.
    To interpret any measure computed on the encapsulation DAG, however, it must be compared against appropriate null models.
    The right panel illustrates the empirical hypergraph alongside the reference hypergraphs used for comparison.
    From left to right, the spectrum of nestedness encompasses the least nested hypergraph comprising the hypergraph's maximal edges, the configuration model as a null model against which to compare empirical measures of nestedness, the empirical network itself, and the induced simplicial complex representing the maximally nested higher-order network.}
    \label{fig:illustration}
\end{figure}

In Fig.~\ref{fig:illustration}, we present our unified framework for quantifying nestedness. The framework centers on the encapsulation DAG~\cite{larock_encapsulation_2023} as a common representation, expressing each measure of nestedness as a function of the statistics of this graph. To contextualize empirical observations, these statistics are compared against those of a null model. In Refs.~\cite{larock_encapsulation_2023,landry_simpliciality_2024}, the null model is the induced simplicial complex---the maximally nested hypergraph sharing the same maximal edges. In contrast, Ref.~\cite{barrett_counting_2025} uses the hypergraph configuration model~\cite{chodrow_configuration_2020} as a randomized baseline. For both the empirical hypergraph and the null model, we compute the statistics of their associated encapsulation DAGs and compare them.

The number of nodes and edges in the encapsulation DAG provide useful information about the nestedness of a system because the number of nodes in the DAG corresponds to the number of edges in the original hypergraph and the number of edges in the DAG corresponds to the number of subset relationships present in the hypergraph.
The in-degree $\din(e)$ of a node $e \in \graph$ counts the number of hyperedges that contain $e$, and the out-degree $\dout(e)$ counts the number of hyperedges that are subsets of $e$.
Because we solely consider simple hypergraphs, all edge subsets must also be proper edge subsets.
The set of maximal hyperedges $\Etilde$, i.e., hyperedges not contained by any other hyperedge, correspond to the nodes with zero in-degree, i.e., $\din(e)=0$.
A hyperedge is a simplex if its out-degree is equal to the size of its power set minus two; one accounting for the edge itself and one accounting for the empty set, i.e., $\dout(e) = |\mathcal{P}(e)|-2$.

The encapsulation DAG is closely related to a Hasse Diagram representing a partial ordering of a set of sets.
However, a Hasse Diagram is transitively reduced by construction, meaning that an edge between two nodes is removed if there is an alternative path between the nodes.
Hasse Diagrams with weights associated to their nodes have been used to define weighted simplicial complexes of hypergraphs, which were further used to predict the evolution and recurrence of small groups \cite{sharma_weighted_2017,sharma_hypergraph_2020}.
We will see later that transitive reduction of an encapsulation DAG into a Hasse Diagram can be used to understand structural patterns of nestedness through studying shortest path lengths.

In Section~\ref{section:measures}, we show that many existing measures of nestedness can be expressed as statistics of the encapsulation DAG, typically compared against the corresponding statistic of a null model. 
As mentioned briefly above, we focus on two specific null models that are useful for defining measures of nestedness.
First is the \textit{induced simplicial complex}~\cite{landry_simpliciality_2024}, formed by adding all hyperedges needed to satisfy downward closure.
We denote this deterministic null model as $\HSC$ with its corresponding encapsulation DAG $\GSC=(\ESC, \DSC)$.
Similarly, we define $\hypergraph^0$ as the hypergraph formed by removing all of the hyperedges in $\hypergraph$ which are included by other edges.
Second is the hypergraph configuration model, a random generative network model which is fully specified by a degree and edge size sequence pair.
This model specifies a distribution over all hypergraph realizations with the same degree and edge size distributions and is most commonly sampled using Markov chain Monte Carlo (MCMC) methods, where edges are shuffled to randomize the hypergraph while preserving these sequences~\cite{chodrow_configuration_2020}.
This model is quite popular because it is straightforward to fit it to data; one simply starts with the original hypergraph and shuffles the edges until the hypergraph is sufficiently randomized, ensuring that the sampled network will have the same degree and edge size sequences, but otherwise random structure.
We denote a single realization of the hypergraph configuration model as $\HCM$ with corresponding encapsulation DAG $\GCM=(\ECM, \DCM)$.

\section{Measures of nestedness}
\label{section:measures}
We now describe a suite of measures for characterizing the nested structure of a hypergraph. These measures fall into two broad categories: \emph{local} measures, which quantify nestedness relative to individual edges or node neighborhoods, and \emph{global} measures, which quantify nestedness for the entire hypergraph.
Several desirable properties have been identified for nestedness measures \cite{landry_simpliciality_2024}. First, they should be undefined on empty hypergraphs, since nestedness requires at least one hyperedge. Second, nestedness spans a spectrum between two extremes: a hypergraph composed entirely of maximal hyperedges (minimal nestedness) and a simplicial complex (maximal nestedness). Third, global measures should be normalizable to the interval $[0,1]$, where 0 corresponds to no inclusion relationships and 1 to a simplicial complex. Fourth, adding inclusion relationships to a hypergraph, all else being equal, should increase the measured nestedness.

In this Chapter, we focus on three recent studies presenting novel ways of measuring nestedness.
First, we describe simple properties of the encapsulation DAG presented in Ref.~\cite{larock_encapsulation_2023} that can characterize nested structure, such as its out-degree distribution.
Second, we describe the measures of \emph{simpliciality} defined in Ref.~\cite{landry_simpliciality_2024} as properties of the encapsulation DAG.
Simpliciality, broadly defined, measures how similar a hypergraph's inclusion structure is to a simplicial complex~\cite{landry_simpliciality_2024}. 
Ref.~\cite{landry_simpliciality_2024} introduces three different measures of simpliciality: the simplicial fraction, the edit simpliciality, and the mean face simpliciality.
These measures highlight different structural elements contributing to the inclusion structure.
Third, we describe the measures introduced in Ref.~\cite{barrett_counting_2025} in terms of the encapsulation DAG.
Two measures of nested structure were introduced in this work: the \textit{simplicial ratio} and the \textit{simplicial matrix}.
Both concepts are built around the notion of the \textit{simplicial pair}, which is equivalent to encapsulation: two edges $e_1$ and $e_2$ are a simplicial pair if $e_1 \subset e_2$.

\subsection{Global Nestedness}
As mentioned previously, by constructing the encapsulation DAG, we can immediately count the total number of subset relationships, which is the number of edges in the DAG.
This might be the simplest global measure of nested hypergraph structure; however, it has limited utility since the absolute number of edges alone does not allow for meaningful comparison between hypergraphs with different sizes.
In this section, we describe three measures of nestedness based on the notion of simpliciality, or distance from maximal nestedness \cite{landry_simpliciality_2024, barrett_counting_2025}.
We describe the intuition behind each measure as well as how they can be computed from the encapsulation DAG.

\subsubsection{Simplicial Fraction}
In a simplicial complex, every subface is itself a simplex, so when a hypergraph is a simplicial complex, it contains all subsets of each of its hyperedges. 
The \emph{simplicial fraction} (SF) measures the degree to which this is true, defined as the fraction of hyperedges which are also simplices. 

Given our hypergraph $\hypergraph = (V, E)$, we let $S = \{ e\in E \mid \ps{e}\subseteq E\}$ be the set of hyperedges which are also simplices. Then, the simplicial fraction is defined as 
\begin{equation}
\ssf = \frac{|S|}{|E|}
\end{equation}
and it takes values in the range $\ssf\in[0,1]$.

The simplicial fraction directly measures the number of simplices in the dataset and is therefore highly interpretable.
However, one potential downside is that edges which \emph{almost} achieve downward closure do not count at all toward the overall simpliciality.
Furthermore, this definition weighs smaller simplices heavily, as small simplices contribute to the simpliciality of all hyperedges that include them.

We can explicitly map this measure to a measure on the encapsulation DAG.
Recall that a hyperedge $e$ is a simplex if all of its proper subsets are also in $E$. This condition is equivalent to the number of outgoing edges from $e$ in $G$ being equal to the number of outgoing edges from $e$ in $\GSC$, i.e., $\dout(e) = \dout_{SC}(e)$. Then, the SF is defined as

\[
\sigma_{\text{SF}} = \frac{\left|\{ e \in E : \dout(e) = \dout_{SC}(e)\}\right|}{|E|}.
\]

\subsubsection{Edit Simpliciality}
The \emph{edit simpliciality} (ES) is defined as the minimal number (or fraction, in the normalized case) of additional edges needed to make a hypergraph a simplicial complex. 

Our formal definition uses the notion of an induced simplicial complex.
Given a hypergraph $\hypergraph = (V, E)$ for which we want to measure the ES, we find its maximal edges $\widetilde{E}$ and construct the simplicial complex $\bS = (V, C)$ induced on $\hypergraph$, with $C=\cup_{e \in \widetilde{E}} \ps{e}$. 
The edit simpliciality is then 
\begin{equation}
  \label{eq:ES}
  \ses = \frac{|E| - |\widetilde{E}|}{|C| - |\widetilde{E}|},
\end{equation}
again satisfying $\ses\in[0,1]$.
We subtract the number of maximal edges from the total number of edges so that the edit simpliciality is zero in the case of a hypergraph with no subfaces.
(We note that one can use the induced simplicial complex to define variants of the ES, e.g., a simplicial edit distance $\des = |C| - |E|$ or a normalized distance $\dnes = (|C| - |E|)/(|C| - |\widetilde{E}|) = 1 - (|E| - |\widetilde{E}|)/(|C| - |\widetilde{E}|) = 1 - \ses$.)

The ES answers a complementary question to the SF since it counts missing hyperedges that would make the dataset into a simplicial complex, rather than the fraction of edges that satisfy downward closure.
However, the ES has the disadvantage of being sensitive to outliers, as a handful of large hyperedges with few inclusions will rapidly drive $\ses$ towards $0$.
Indeed, a hyperedge of size $m$ without any inclusion contributes one edge to $|E|$ but $2^m$ edges to $|C|$ in the denominator of Eq.~\eqref{eq:ES}. 

ES quantifies how many missing hyperedges are required to form a minimal simplicial complex induced by the maximal hyperedges $\Etilde$. Recall that the set of maximal hyperedges can be characterized by $\Etilde = \{e \in E: d^{\text{in}}(e)=0\}$. Then the ES is defined as

\[
\sigma_{\text{ES}} = \frac{|E| - |\Etilde|}{|\ESC| - |\Etilde|}.
\]

This expression compares the number of existing non-maximal hyperedges to the number of unique subfaces required to form the induced simplicial complex.

\subsubsection{Simplicial Ratio and Simplicial Matrix}

The \textit{simplicial ratio} measures the number of inclusion relationships, referred to as \textit{simplicial pairs}, in an empirical hypergraph relative to the expected number of simplicial pairs for a configuration model (or equivalently, a Chung-Lu model) hypergraph with the same degree and edge size sequences \cite{barrett_counting_2025}.
As with many of the other measures described, this null model need not be the configuration model, but could be any random or deterministic model for which a comparison makes sense.
Then the simplicial ratio is defined as
\begin{align}
\sr{\hypergraph} & = \frac{\simppair{\hypergraph}}{E[\simppair{\widehat{\hypergraph}}]},
\end{align}
where $\simppair{\hypergraph} = |\{e_1 \subset e_2 \mid e_1, e_2 \in E\}|$ and $\widehat{\hypergraph}$ denotes a random null model.
This expectation is computed by averaging the number of simplicial pairs obtained for each realization in an ensemble of configuration model hypergraphs.
These realizations are sampled from the configuration model by performing random edge shuffles on the empirical hypergraph in line with the MCMC methods described in Ref.~\cite{chodrow_configuration_2020}.

Ref.~\cite{barrett_counting_2025} also introduced an associated measure called the \textit{simplicial matrix}.
This measure provides a more granular look at the nested structure of a hypergraph, now computing the simplicial ratio for each combination of edge sizes.
Specifically, the authors extend the definition of simplicial pairs and introduce $\simppair{\hypergraph, i, j}$, counting the number of times that edges of size $j$ include edges of size $i$, which they denote $\simppair{\hypergraph, i, j}$.
More formally, $\simppair{\hypergraph, i, j} = |\{ e_1 \subset e_2 \mid e_1,e_2 \in E, |e_1|=i, |e_2|=j\}|$.
Then, the simplicial matrix is defined as
\begin{equation}
\msr{\hypergraph, i, j} = \frac{\simppair{\hypergraph, i, j}}{E[\simppair{\hypergraph, i, j}]}.
\end{equation}
The simplicial matrix "unpacks" the simplicial ratio~\cite{barrett_counting_2025}, quantifying the level of nestedness between edges of every size.

We can express both the simplicial ratio and the simplicial matrix in terms of the encapsulation DAG. Recalling that $D$ and $\DCM$ are the edge sets of the encapsulation DAG for the hypergraph $\hypergraph$ and the $\HCM$, respectively, we define $D_{ij}$ and $\DCM_{ij}$ to be the edge sets representing inclusion relationships between edges of sizes $i$ and $j$. Then, the simplicial ratio can be expressed as
\begin{equation}
\sr{\hypergraph} = \frac{|D|}{\langle |\widehat{D}|\rangle},
\end{equation}
where $\widehat{D}$ is the edge set corresponding to a sample from the hypergraph configuration model. Similarly, the $(i,j)$th entry of the simplicial matrix is defined as

\begin{equation}
\msr{\hypergraph, i, j} = \frac{|D_{ij}|}{\langle |\widehat{D}_{ij}|\rangle}.
\end{equation}

\subsection{Local Nestedness}
We now turn to definitions of nestedness that are \emph{local} in the sense that they measure the nested structure of individual hyperedges or nodal neighborhoods, rather than quantifying nestedness of an entire hypergraph.
Local measures allow us to quantify how nested structure is distributed across a hypergraph.
Just as measures such as clustering coefficient can be framed both locally as the density of a node's neighborhood, and a global measure as the average number of closed triangles, we do the same for measures of nestedness.

\subsubsection{Degree distributions in the encapsulation DAG}

As previously discussed, the out-degree of a hyperedge in the encapsulation DAG counts the number of hyperedges nested in that edge, while the in-degree represents the number of hyperedges in which it is nested.
The out-degree of a hyperedge has a well defined maximum based on the size of the hyperedge's power set, which depends on the size of the hyperedge itself.
We can use the distribution of the normalized out-degree to understand how variable local nestedness is among the hyperedges in the hypergraph.
A minimally nested hypergraph would have values concentrated at 0, while a maximally nested hypergraph would have values of normalized out-degree concentrated around the maximum value of 1.

Similarly, the in-degree of a hyperedge in the DAG indicates the extent to which the supersets of a hyperedge exist, e.g., how much a given hyperedge is encapsulated.

\subsubsection{Path lengths in the encapsulation DAG}

The length of paths in the encapsulation DAG can be interpreted as the ``breadth'' or ``depth'' of nestedness relationships.
In particular, it is useful to analyze the height of \textit{rooted paths in the transitively reduced DAG}, an approach inspired by that of Ref.~\cite{vasiliauskaite_cycle_2022}.
A rooted path is one that begins from a root node, defined as a maximal hyperedge or equivalently a node in the DAG with zero in-degree (and non-zero out-degree).
We consider paths starting from root nodes because they indicate the maximum possible path lengths through the encapsulation DAG.
A transitively reduced DAG is one in which all edges representing shorter redundant paths are removed.
For example, if we have the edges A-B, B-C, and A-C, in the transitively reduced DAG the edge A-C would be removed, since there would still be a path from A to C without that edge.
Analyzing the DAG after removing these ``shortcut'' edges gives us a sense for the extent to which intermediate sized hyperedges are or are not present.

The distribution of path lengths in the transitively reduced DAG indicates the depth of the encapsulation relationships in the hypergraph.
If the distribution is skewed towards the maximum length ($k-1$ edges for a hyperedge on $k$ nodes), this indicates a  hierarchy of encapsulations in the sense that  multiple intermediate hyperedges of different sizes are all encapsulated by the same larger hyperedge (the root).
In contrast, if most path lengths are short, this indicates that encapsulation relationships in the hypergraph are concentrated between only two different sizes at a time, a kind of shallow encapsulation.
Note that transitively reduced DAGs corresponding to two hypergraphs with very different encapsulation structures could have similar numbers of edges, but very different path length distributions.
Deeper and more hierarchical encapsulation relationships can have important implications for how a contagion can spread over the hyperedges of a hypergraph.

\subsubsection{Face Edit Simpliciality}

Building upon the idea of the global edit simpliciality defined above, a more localized notion of simpliciality can also be formulated using the number of subfaces that must be added to the hypergraph to make a particular face a simplex.

Given a hyperedge $e$, the number of edges one must add to the hypergraph to make $e$ a simplex is 
\begin{equation*}
  d_{\mathrm{FES}}(e) = |\ps{e}| - |c|,  
\end{equation*}
where $c = \{ f \in E \mid f \subseteq e\}$.
We can think of this quantity as an edit distance, or \emph{face edit distance}.
We use this quantity to define an average 
\begin{equation*}
  \bar{d}_{\mathrm{FES}} = \frac{1}{\left| F\right|}\sum_{e\in F} d_{\mathrm{FES}}(e),
\end{equation*}
where $F$ is a set of edges---most commonly, $F = \widetilde{E}$ or $E$.
Here, we assume that $F = \widetilde{E}$.
These quantities are on the scale of counts, and to define quantities analogous to previous simpliciality measures, we thus introduce a per-face normalization, either on a distance scale (meaning that the quantity grows as the dataset becomes less simplicial):
\begin{equation*}
  \bar{d}_{\mathrm{NFES}} = \frac{1}{\left| F\right|}\sum_{e\in F} \frac{d_{\mathrm{FES}}(e)}{\left|\ps{e}\right|-1},
\end{equation*}
or, similarly to previous definitions, on a simpliciality scale:
\begin{equation}
  \sfes = \frac{1}{\left| F\right|}\sum_{e\in F} \left(1 - \frac{d_{\mathrm{FES}}(e)}{\left|\ps{e}\right|-1}\right).
\end{equation}
We call this last measure the \emph{face edit simpliciality} (FES), which, in its averaged form is actually a global measure.
We subtract one in the denominators of both expressions so that when an edge has no subfaces, its normalized face edit distance is one.

The FES normalizes the face edit distance as a fraction of its maximal simpliciality. 
This normalization removes the dominance of large edges in the calculation of $\ses$ and, in fact, exponentially down-weights the contribution of these edges. 
In addition, because this metric is computed on faces, this is an averaged local metric. 

Recalling that $\Etilde$ denotes the maximal edges and that the out-degree of an edge is the number of edges contained by that edge, the FES can be represented in terms of the encapsulation DAG as
\[
\sigma_{\text{FES}} = \frac{1}{|\Etilde|} \sum_{e \in \Etilde} \frac{\dout(e)}{\dout_{\text{SC}}(e)},
\]
where $\dout_{SC}(e) = |\mathcal{P}(e)| - 2$.

\subsubsection{Nodal Simpliciality}

Any measure of nestedness defined on a hypergraph can also be localized on a smaller subset of the higher-order network to yield information about its local structure.
Here, we use subhypergraphs to define \emph{nodal simpliciality} measures of our global measures.
More specifically, given a hypergraph $\hypergraph = (V,E)$ and a node $v\in V$, we define the neighborhood of $v$ as $n(v) = \{u \in V \mid u, v \in e\in E\}$ and the associated subsets $\widehat{V}=v\,\cup\, n(v)$ and $\widehat{E} = \{e \in E \mid e\subseteq \widehat{V}\}$. Then the nestedness of node $v$ is the nestedness defined on the subhypergraph $\widehat{\hypergraph} =(\widehat{V}, \widehat{E})$ induced on the neighborhood of $v$. Note that when $v$ is an isolated node or when $\widehat{E}$ only contains minimal faces and we do not consider these potential simplices, the nodal nestedness will be undefined.


\subsection{Considerations}
\label{sec:limits}

There are three design choices that may impact the conclusion we reach about the nestedness of a dataset, as discussed in Ref.~\cite{landry_simpliciality_2024}.

First, the formal definition of a simplicial complex can be unnecessarily strict when used to represent perfect inclusion structures.
By definition, a simplex always contains singletons (edges comprising a single node) and the empty set. 
Several datasets will not include such interactions by construction.
One example is proximity datasets, where edges encode proximity events in which two or more nodes become in close contact during the observation period.
Because of their spatial nature, these datasets are often very dense and contain many inclusions~\cite{battiston_networks_2020}.
Yet, according to the standard definition, these will never be simplicial complexes due to the absence of singletons.
Another example is email datasets, which also do not contain singletons unless one includes emails that individuals send to themselves.
Because we define our notion of inclusion in terms of simplicial complexes, our measures will label these datasets as having no inclusion structure whatsoever.

To circumvent this issue, we use a relaxed definition of downward closure that excludes singletons wherever it makes sense.
The relaxation uses the notion of a \emph{size-restricted power set} $\srps{X}{K}$, where $K$ is a set of integers, defined as
\begin{equation}
  \label{eq:size_restricted_power_set}
  \srps{X}{K} = \left\{ x \in \ps{X} \ \Big| \ |x| \in K\right\}.
\end{equation}
For example, given an edge $e$ of size $n$, $\srps{e}{\{2,\dots,n-1\}}$ is the set of $2^{|e|} - |e| - 2$ subfaces of $e$ excluding the empty set, all singletons (sets of size one), and the edge $e$ itself.
Relaxed measures of nestedness follow by substituting $\ps{X}$ for $\srps{X}{K}$ in all nestedness measures.
Hence, for example, we obtain a relaxation of $\ssf$ by replacing the definition of $S$, the set of the hyperedges of $\hypergraph$ that are also simplices, by $S = \{e\subseteq E\mid\srps{e}{K}\subseteq E\} $), where $K=\big\{2,...,|e|\big\}$.

We calculate results using size restrictions to exclude singletons and the empty set.
However, we note that this technique can be used more generally to exclude any interaction sizes deemed unimportant, anomalous, or problematic \cite{landry_filtering_2024};  or, conversely, to be more strict and to include singletons (say, when analyzing academic co-authorship networks, where single-author papers can meaningfully impact the inclusion structure of the dataset).

Second, we observe that special hyperedges we call ``minimal faces'' may significantly skew the simpliciality of a dataset.
The \emph{minimal faces} of a hypergraph $\hypergraph$ are the edges of the minimal size, i.e., $|e| = \min(K)$, where $K$ is the set of sizes in the size-restricted powerset (In a traditional simplicial complex, the minimal faces are singletons).
With the size restrictions in place, the minimal faces of a hypergraph are always simplices because, by definition, there are no smaller edges for these edges to include.
We argue that when measuring the simpliciality of a dataset, it is most meaningful to focus on the faces for which inclusion is \emph{possible}, and so we exclude these minimal faces when counting potential simplices.

Note that this design choice operates differently from the size restriction imposed by the modified power set introduced in Eq.~\eqref{eq:size_restricted_power_set}; in that context, we argued for ignoring edges that can prevent other edges from being simplices, while here we suggest that counting minimal faces as potential simplices will strongly affect the value of nestedness.

Third and finally, since the number of potential subfaces of a hyperedge grows exponentially with its size, computational issues prevent us from applying our measures to large hyperedges.
For this reason, we select a maximum face size $k$ (we use $k=11$ throughout), again using the size restriction to define our metrics.
This drops information about large hyperedges but speeds up computation drastically in practical applications.
To avoid these computational issues, one should avoid reconstructing the induced simplicial complex whenever possible.
For many measures, it is easy to avoid this, because one can simply calculate the size of the power set as $2^{|e|}$ or the size-restricted power set as $\sum_{k\in K} \binom{|e|}{k}$.
One exception to this is the edit simpliciality.
While this measure is explainable, it can be quite challenging to calculate in a memory-efficient manner.
While Ref.~\cite{landry_simpliciality_2024} introduces an algorithm that is more efficient than constructing the induced simplicial complex, it still requires constructing parts of the induced simplicial complex.

\subsection{Constructing the encapsulation DAG}
\label{subsec:construct-dag}
A challenge in studying nestedness in hypergraphs is the potential for combinatorial explosion in the number of comparisons needed, since for a hyperedge of size \(\alpha\), the number of possible subsets may be as large as \(2^\alpha - 2\), corresponding to the power set of the hyperedge minus the set itself and the empty set.
Therefore it is worthwhile to consider how the encapsulation DAG can be constructed and then used to efficiently measure nested structure.

Here we describe a straightforward algorithm to construct an encapsulation DAG.
We give pseudocode for this procedure in Algorithm~\ref{alg:DAG}.
We first assign each hyperedge $e$ a unique label $\texttt{map}_e \in 1,\dots,|E|$ and construct a node-membership lookup table $\texttt{memb}_u$ between each node and the list of hyperedges it participates in (lines 2-7).
We then loop over each hyperedge $\alpha \in E$, and for each node $u\in \alpha$ we add edges from $\alpha$ to other hyperedges $\beta \in \texttt{memb}_u$ in which $u$ participates if $\beta$ is encapsulated by $\alpha$, meaning we add edges where $\beta \subset \alpha$ (or {\it vice-versa}).(lines 8-12).
After repeating this loop for each node in $\alpha$, the out-neighbors of $\alpha$ in $L$ represent all of the hyperedges $\beta \in E$ that are subsets of $\alpha$.

\begin{algorithm}
	\caption{Construct an encapsulation DAG for a hypergraph $\hypergraph=(V, E)$.}
	\label{alg:DAG}
	\begin{algorithmic}[1]
    \Procedure{EncapsulationDAG}{$\hypergraph$}
        \State $\texttt{memb}_u \gets \emptyset$
        \For{$ i \in 1,\cdots,|E|$}
            \For{$u \in E_i$}
                \State $\texttt{memb}_u \gets \texttt{memb}_u \cup i$
            \EndFor
        \EndFor
        \State $L \gets \emptyset$
        \For{$\alpha \in E$}
            \For{$u \in \alpha$}
                \For{$\beta \in \texttt{memb}_u$}
                    \If{$\beta \subset \alpha$}
                        \State $L \cup (\alpha \to \beta)$
                    \ElsIf{$\alpha \subset \beta$}
                        \State $L \cup (\beta \to \alpha)$
                    \EndIf
                \EndFor
            \EndFor
        \EndFor
    \EndProcedure
    \Return L
	\end{algorithmic}
\end{algorithm}

The complexity of this construction has two terms corresponding to the two nested for loops.
The first doubly nested loop runs over all  $m=|E|$ hyperedges to construct a mapping from hyperedges to labels, then all of the nodes in each hyperedge to fill the mapping from nodes to the hyperedges in which they appear.
This operation takes $O(m\cdot s_{\max})$ time, where $s_{\max} = \max_{e \in E}{s_e}$ is the maximum size of a hyperedge.
Once the mappings are constructed, we enter the triply nested loop to find encapsulation relationships.
The worst case time for an inner loop is the size of the largest hyperedge $s_{\max}$ times the highest degree node $k_{\max} = \max_{u \in V} |\{e | u \in e; e\in E\}|$. Clearly the triply nested loop dominates the doubly nested loop and so the worst case running time is $O(m\cdot s_{\max}\cdot k_{\max})$.

\section{Generative models}\label{sec:generative}

To our knowledge, there are only a couple of generative higher-order network models that explicitly parameterize nestedness, and these models are introduced in Refs.~\cite{kim_contagion_2023,barrett_counting_2025}.
While, for example, Ref.~\cite{zhang_higherorder_2023} reduced the inclusion structure by performing random edge shuffling, they did not explicitly specify the nestedness; rather, it was an outcome of this hypergraph modification process.

Reference \cite{kim_contagion_2023} introduces a random nested-hypergraph model.
The model generation process comprises three steps and is parametrized by three fixed quantities: the size of facets $D$ (a reference to simplicial complexes), the number of facets $M$, and the number of nodes $N$. 
First, one generates $M$ facets of size $D$ and assigns nodes to each facet uniformly at random; nodes may belong to multiple facets at this stage, but facets that overlap completely are rejected. Next, the hyperedges are created: One for each facet and all the proper subsets of these facets, excluding sets of cardinality one. The resulting hypergraph corresponds to a classical simplicial complex with its inclusion property.
Nestedness is finally tuned by rewiring the hyperedges of size less than $D$. More precisely, a hyperedge of size $\alpha<D$ is rewired with probability $1-\epsilon_\alpha$ by selecting a pivot node uniformly at random from the hyperedge and replacing the other nodes with randomly sampled nodes. 

\begin{figure}
    \centering
    \includegraphics[width=\linewidth]{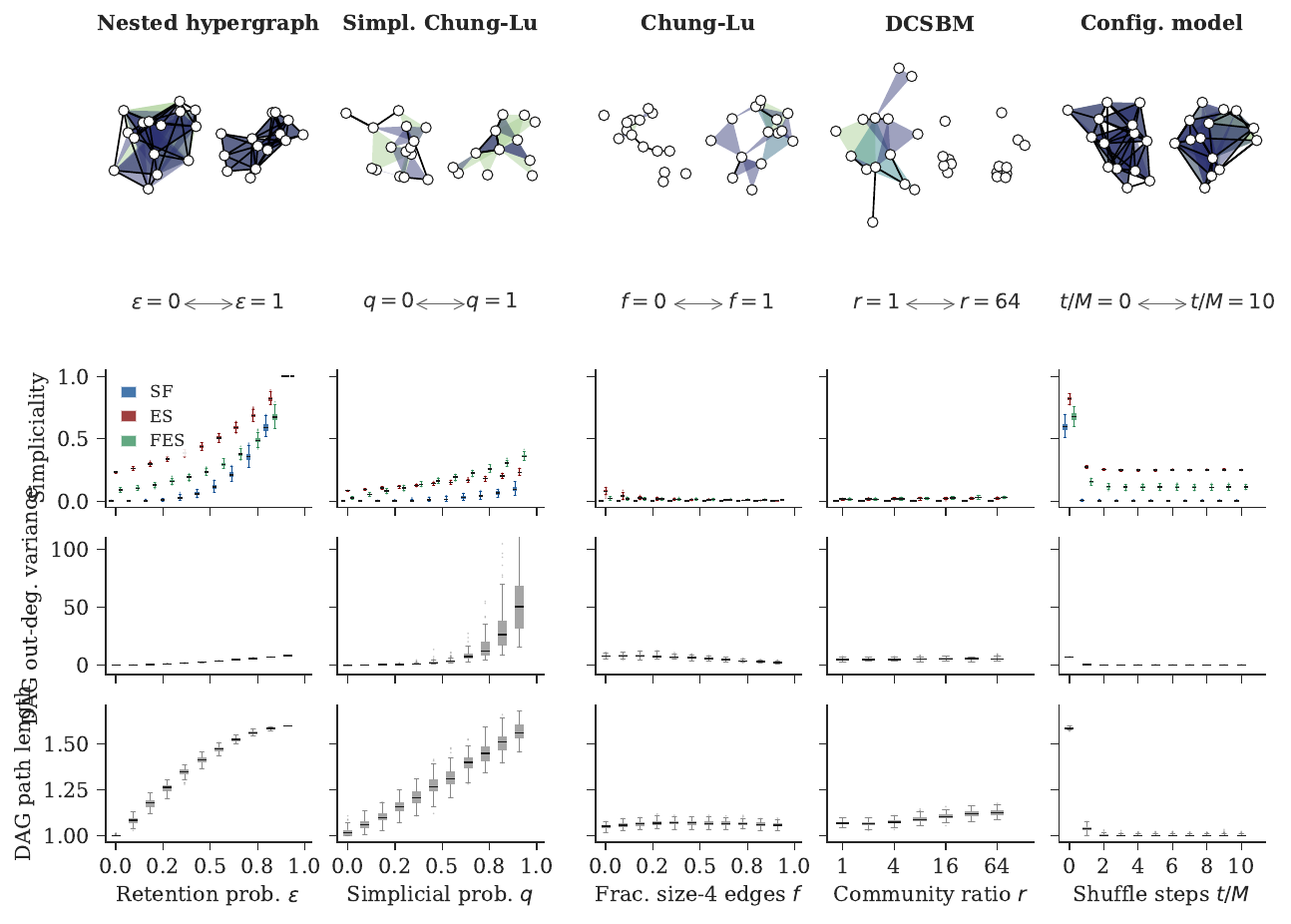}
    \caption{\textbf{Nestedness in random models of hypergraphs.} The first two columns show two models that explicitly control for the nestedness of the hypergraphs, while the last three columns show models capturing different properties, like the degree sequence of the community structure. Example hypergraphs are shown at the top (with fewer nodes for clarity), while the subsequent rows plot the various nestedness statistics described in Sections~\ref{section:measures}; the first row plots the simplicial fraction (SF), the edit simpliciality (ES), and the face edit simpliciality (FES); the second row plots the out-degree variance of the encapsulation DAG; and the last row plots the average path length in the DAG. Quantities are calculated from 100 sampled hypergraphs ($N=100$ nodes) for each model and parameter value, with one key parameter varied across a representative range. Fixed parameters are set to:  $M=75$ initial facets of size $D=4$ for the random nested hypergraph model described in Ref.~\cite{kim_contagion_2023}; 500 edges divided equally between sizes $\{2,3,4\}$ for the simplicial Chung-Lu model introduced in Ref.~\cite{barrett_counting_2025}; 500 edges divided equally between sizes $\{2,4\}$ for the hypergraph Chung-Lu model~\cite{aksoy_measuring_2017}; 500 edges with sizes in $\{2,3,4\}$ in the same proportion as the simplicial Chung-Lu experiments, and two equally-sized communities for the hypergraph DCSBM~\cite{yen_community_2020}; swaps applied to a random initial hypergraph generated with the nested hypergraph model and parameters $\varepsilon=0.9$, $M=75$, $D=4$ to sample from the hypergraph configuration model~\cite{chodrow_configuration_2020}, where a ``shuffle step'' refers to a sequence of $|E|$ random edge shuffles as described in Ref.~\cite{chodrow_configuration_2020}.}
    \label{fig:null_models}
\end{figure}

A different approach is taken in Ref.~\cite{barrett_counting_2025}, which introduces a modification of the hypergraph Chung-Lu model where, with probability $1-q$, edges are sampled according to the Chung-Lu model; i.e., edges are selected with probability $p=(\alpha-1)! k_1\dots k_{i_\alpha}/(N\langle k\rangle_\alpha)^{\alpha-1}$.
To control the nestedness of the hypergraph, new edges are sampled with probability $q$ as subsets of existing edges.
The paper demonstrates that (1) the nestedness of the hypergraph increases with $q$ and (2) that the degrees are matched in expectation and the edge size sequences are matched exactly.
This algorithm can be used to match not only the simplicial ratio, but also the simplicial matrix.

As discussed in Ref.~\cite{landry_simpliciality_2024}, current null models, such as the hypergraph configuration model~\cite{chodrow_configuration_2020}, the Chung-Lu model~\cite{aksoy_measuring_2017}, and the degree-corrected stochastic block model~\cite{yen_community_2020}, fail to accurately capture the nestedness of empirical datasets.
To compare and contrast the behavior of these standard null models with that of models designed to produce highly nested structure, we ran the small simulation study shown in Fig.~\ref{fig:null_models}.
The two models on the left control nestedness directly, while the three others do not.
By all measures---from edit simpliciality to encapsulation DAG summaries---the models of Refs.~\cite{kim_contagion_2023,barrett_counting_2025} display much higher nestedness.
The configuration model study is particularly instructive.
In this experiment, we initialized the algorithm with a random nested hypergraph generated with a retention probability of $\epsilon=0.9$ for all hyperedge sizes, resulting in a structure close to a perfect simplicial complex.
After only a few shuffling steps ($t=2$ corresponding to two sweeps of $|E|$ random hyperedge swaps), all traces of nestedness were practically erased.

Thus, while a few null models now exist to represent nestedness, there remains ample opportunity to develop network models that parameterize it.
The basic algorithms described in this study could serve as the foundation for a suite of network models that can be fitted to empirical data.

\section{Case study}\label{sec:case-study}

To illustrate the nested structure of complex social systems, we measured nestedness in several datasets across a variety of domains and use the differences across measures to explain the intrinsic properties of each dataset.

In the Science of Science domain, we selected the \texttt{coauth-mag-geology} dataset~\cite{sinha_overview_2015, benson_simplicial_2018, landry_xgi_2023}, where nodes represent authors and hyperedges represent published papers. We chose this dataset because of the ubiquity of co-authorship networks in the network science literature.
For proximity-based contact networks, we leveraged two datasets: the \texttt{contact-high-school} dataset~\cite{mastrandrea_contact_2015}, where nodes represent individuals tracked via RFID tags and hyperedges record face-to-face proximity events, and the \texttt{malawi-village} dataset~\cite{ozella_using_2021}, which captures close social contacts among residents of a rural Malawian village.
In the political domain, we analyzed the \texttt{senate-bills} and \texttt{house-bills} datasets~\cite{chodrow_generative_2021, fowler_connecting_2006, fowler_legislative_2006, landry_xgi_2023}, where nodes denote U.S. congresspersons, and hyperedges are formed by the co-sponsors of individual bills. 
All datasets were retrieved from XGI-DATA~\cite{landry_xgi_2023}.

Considering the structural limitations mentioned in Section~\ref{sec:limits}, we first preprocessed each dataset by removing singletons, multi-hyperedges and isolated nodes. Also, for computational tractability, we filtered the datasets~\cite{landry_filtering_2024}, only considering hyperedges of size $11$ and smaller. Next, we converted these preprocessed datasets into their corresponding encapsulation DAGs.
Using these DAGs, we computed several simpliciality measures: simplicial fraction $\sigma_{\text{SF}}$, edit simpliciality $\sigma_{\text{ES}}$, face-edit simpliciality $\sigma_{\text{FES}}$, together with the simplicial ratio $\sigma_{\text{SR}}$ (total observed inclusions divided by the expected number under the Chung-Lu model, averaged over $1000$ Monte Carlo samples) and the average path length $\left\langle L_\text{TR}\right\rangle$ of the transitively reduced DAGs. The summary statistics are shown in Table~\ref{tab:properties}.

\begin{table}[ht]
    \centering
    \renewcommand{\arraystretch}{1.3} 
    \begin{tabular*}{\textwidth}{@{\extracolsep{\fill}} l r r r r r r r @{}}
        \toprule
        Dataset & $|V|$ & $|E|$ & $\sigma_{\text{SF}}$ & $\sigma_{\text{ES}}$ & $\sigma_{\text{FES}}$ & $\sigma_{\text{SR}}$ & $\left\langle L_\text{TR}\right\rangle$\\
        \midrule
        
        \textbf{Science of Science}&&&&&&&\\
        \quad \texttt{coauth-mag-geology} & 1,061,562 & 898,649 & 0.00 & 0.01 & 0.05 & 2395.23 & 1.23 \\
        \addlinespace
        
        \textbf{Proximity-based contact}&&&&&&&\\
        \quad \texttt{contact-high-school} & 327 & 7,818 & 0.81 & 0.91 & 0.92 & 6.55 & 1.18 \\
        \quad \texttt{malawi-village} & 86 & 432 & 0.96 & 0.99 & 0.98 & 5.48 & 1.09 \\
        \addlinespace
        
        \textbf{Politics}&&&&&&&\\
        \quad \texttt{senate-bills} & 293 & 16,122 & 0.03 & 0.00 & 0.12 & 1.92 & 1.35 \\
        \quad \texttt{house-bills} & 1491 & 28401 & 0.00 & 0.00 & 0.05 & 4.74 & 1.27 \\
        \bottomrule
    \end{tabular*}
    \caption{Network properties of selected datasets.}
    \label{tab:properties}
\end{table}

\begin{figure}[ht]
    \centering
    \includegraphics[width=\linewidth]{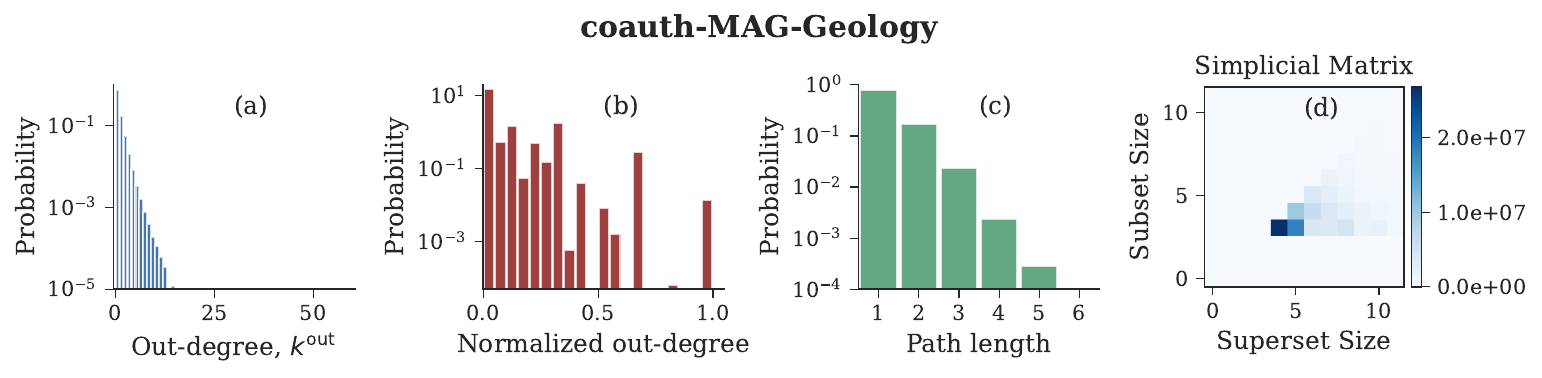}
    \caption{\textbf{Nestedness Statistics for the \texttt{coauth-mag-geology} dataset.} Panel (a) plots the out-degree distribution of the DAG, panel (b) plots the distribution of the out-degree, normalized by the maximum possible value for an edge of that size, panel (c) plots the distribution of path lengths through the transitively reduced DAG (Hasse diagram), and lastly, panel (d) plots the simplicial matrix, where we visualize the entries as a heat map.}
    \label{fig:geology}
\end{figure}

\begin{figure}[ht]
     \centering
     \begin{subfigure}[b]{\textwidth}
         \centering
         \includegraphics[width=\textwidth]{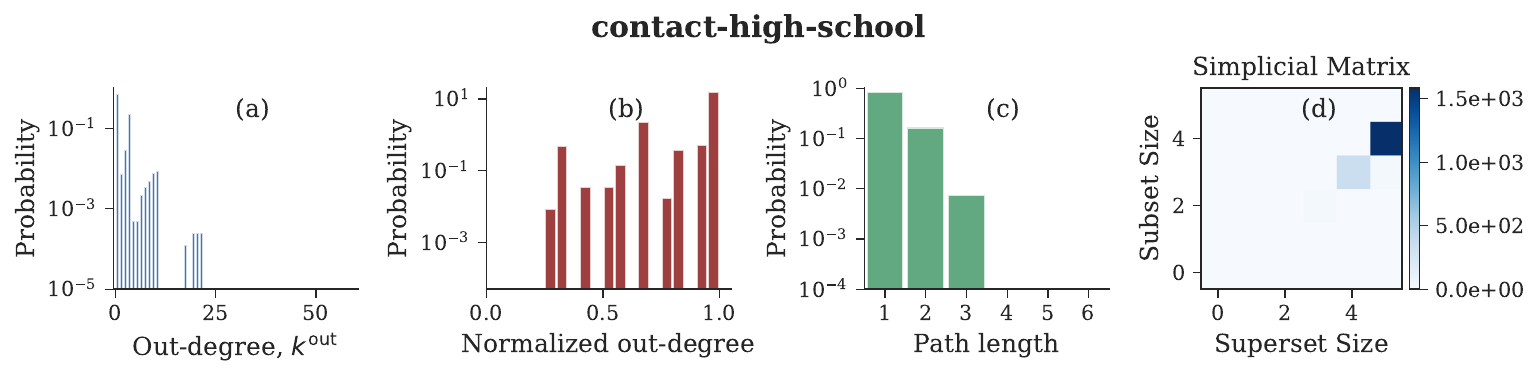}
         \label{fig:contact_left}
     \end{subfigure}

     \begin{subfigure}[b]{\textwidth}
         \centering
         \includegraphics[width=\textwidth]{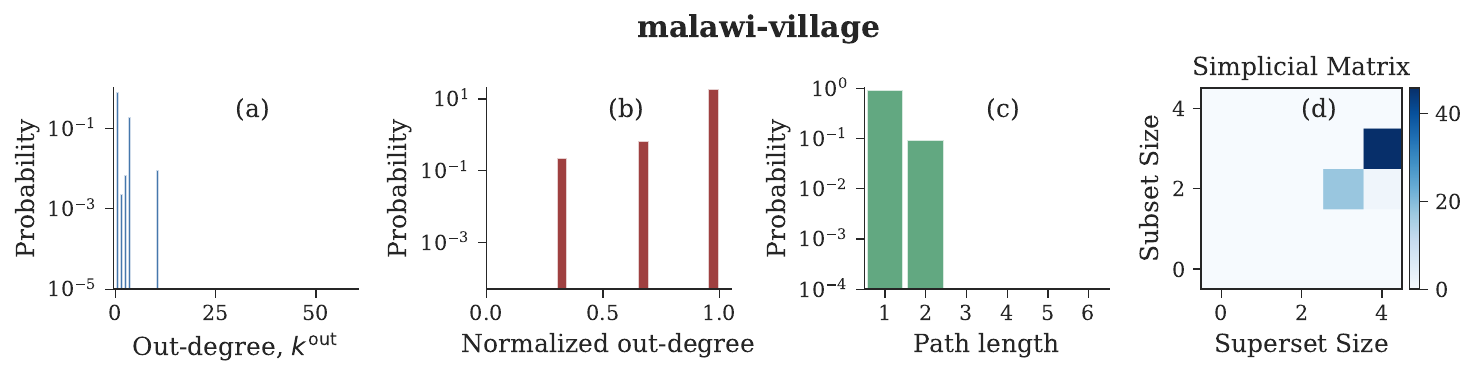}
         \label{fig:contact_right}
     \end{subfigure}
     \caption{\textbf{Nestedness in proximity-based contact networks.} In both plots, panel (a) plots the out-degree distribution of the DAG, panel (b) plots the distribution of the out-degree, normalized by the maximum possible value for an edge of that size, panel (c) plots the distribution of path lengths through the transitively reduced DAG (Hasse diagram), and lastly, panel (d) plots the simplicial matrix, where we visualize the entries as a heat map.}
     \label{fig:proximity}
\end{figure}

\begin{figure}[ht]
     \centering
     \begin{subfigure}[b]{\textwidth}
         \centering
         \includegraphics[width=\textwidth]{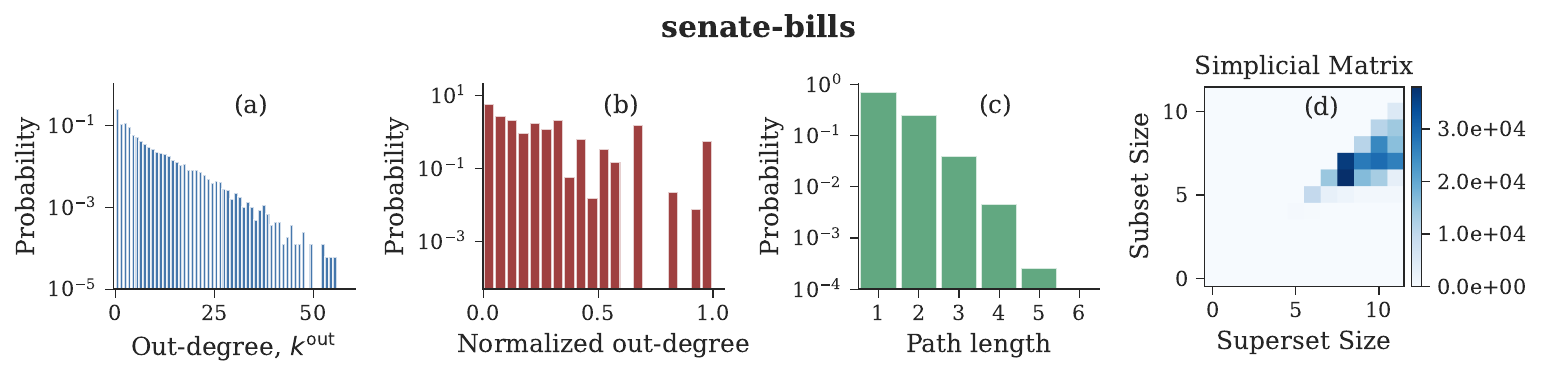}
         \label{fig:politics_left}
     \end{subfigure}

     \begin{subfigure}[b]{\textwidth}
         \centering
         \includegraphics[width=\textwidth]{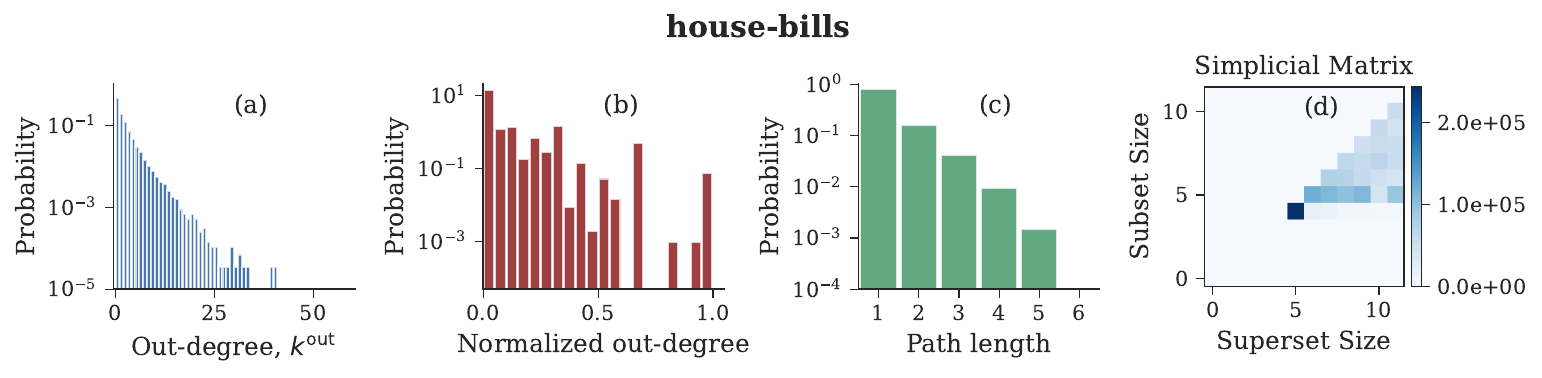}
         \label{fig:politics_right}
     \end{subfigure}
     \caption{\textbf{Nestedness in political co-sponsorship networks.} In both plots, panel (a) plots the out-degree distribution of the DAG, panel (b) plots the distribution of the out-degree, normalized by the maximum possible value for an edge of that size, panel (c) plots the distribution of path lengths through the transitively reduced DAG (Hasse diagram), and lastly, panel (d) plots the simplicial matrix, where we visualize the entries as a heat map.}
     \label{fig:politics}
\end{figure}

Along with these quantitative results, we also provide additional visualizations for each dataset in Figures~\ref{fig:geology}-\ref{fig:politics}. These include the raw and normalized out-degree distributions of the encapsulation DAG, histograms of path lengths (depth), and the simplicial matrix.

From Table~\ref{tab:properties}, we observe that the proximity-based datasets exhibit high levels of global nestedness, with $\sigma_{\text{SF}}$ reaching $0.81$ for the high school and $0.96$ for the village. Both $\sigma_{\text{ES}}$ and $\sigma_{\text{FES}}$ are above $0.9$, which implies very few subsets are missing from the induced simplicial complex.
This near-perfect nestedness is primarily due to the physical constraints of social contact datasets. For example, if a group of people is recorded as interacting within a restricted spatial radius, then it is inevitable that any subset within that group will also be close in proximity, and these subset interactions will be recorded as the group grows, shrinks, and dissolves over time. This will naturally drive the hypergraph toward downward closure and resulting in a highly nested structure.

In contrast, the co-authorship network presents a different structural paradigm, showing a very low level of nestedness where $\sigma_{\text{SF}}$, $\sigma_{\text{ES}}$ and $\sigma_{\text{FES}}$ are all nearly $0$. This sparsity may be due to the near-perfect downward closure requirement of simpliciality metrics. For example, if a team of ten researchers co-authors a geology paper, for this group to contribute to the simplicial fraction, they would need to publish over a thousand independent papers covering every possible smaller combination of those ten individuals, which is practically impossible due to the relatively high cost of scientific collaborations.
However, the simplicial ratio $\sigma_{\text{SR}}$, which measures the number of inclusions relative to the expected number of simplicial pairs for a configuration model, is notably higher than all other datasets at $2,395.23$. Furthermore, the majority entries of its simplicial matrix are large, at more than $1,000$. This indicates that while perfect downward closure is globally rare, subset inclusions actually occur at a rate much higher than random chance, likely reflecting subgroups representing collaborations within and across different research groups. This apparent topological paradox highlights the necessity of using multiple nestedness metrics when analyzing complex group interaction datasets.

Finally, we examine political networks through the senate-bills and house-bills datasets. Similar to the scientific collaboration network, political co-sponsorship yields a very low $\sigma_{\text{SF}}$ and $\sigma_{\text{ES}}$. However, we note that the average transitively reduced DAG path lengths $\langle L_{\text{TR}} \rangle$ for the Senate ($1.35$) and House ($1.27$) are the highest among all observed domains. Also, the distributions of path lengths are skewed towards the maximum length of $5$ and $6$, implying a clear hierarchy of nestedness where multiple intermediate hyperedges of different sizes are encapsulated by the same larger hyperedges. In a legislative setting, a highly specific bill co-sponsored by a small group of legislators may be encapsulated by another bill co-sponsored by a larger, mid-sized group, which is itself fully contained within a broader piece of legislation co-sponsored by an even larger group. This deep hierarchy reflects the underlying organization of the legislative system, from general legislative matters relevant to all (\emph{e.g.}, budgets), to delegated bureaucratic units (\emph{i.e.}, committees), to highly specific pieces of legislation (\emph{e.g.}, pertaining to localized geographic regions).

\section{Discussion}

In this Chapter, we reviewed several measures of nestedness and unified them by expressing each as a function of the encapsulation DAG. This common formulation clarifies the functional differences between measures and facilitates direct comparison. Nestedness is inherently a higher-order phenomenon, and quantifying it provides insight into the formation, structure, and dynamics of complex systems that is often inaccessible through traditional network methods.

While we synthesize and unify the state-of-the-art analytical methods and measures, we remark that there remain a vast number of unanswered questions.
Chief among them is, what are the mechanisms that lead to the highly nested structure that we observe in nature?
Ref.~\cite{iacopini_temporal_2024} introduces a mechanistic null model that captures the temporal evolution of groups, but does not explicitly capture the nested structure of empirical datasets, though this mechanism may well capture much of the inclusions formed by groups merging and splitting.
Similarly, though Ref.~\cite{barrett_counting_2025} described a null model which explicitly parameterizes nestedness, it does not fully answer the question: how does nestedness emerge in empirical systems?
While this modified Chung-Lu model is able to match not only the degree distribution but the inclusion structure as well, it does not capture plausible ways in which a network can grow to exhibit these properties.
New models explaining the evolution of nestedness from a network growth perspective would complement foundational work proposing plausible mechanisms to explain heterogeneous degree distributions~\cite{barabasi_emergence_1999,overgoor_choosing_2019}, for example.
Future work could include the development of mechanistic and descriptive models in each of these areas.
While Ref.~\cite{landry_simpliciality_2024} provided evidence that common null models fail to capture nestedness in empirical systems, this study was by no means exhaustive.
It would be fruitful to more systematically examine a larger set of null models to identify network models which plausibly reproduce nestedness patterns in empirical data.

A major limitation of existing nestedness measures is the inability to handle temporal nestedness, multihyperedges, or weighted hyperedges. Many empirical datasets naturally carry richer information than we have studied here: proximity datasets from projects such as SocioPatterns contain repeated interactions, co-authorship datasets feature multiedges, and in other settings weights may encode interaction intensities or edge existence probabilities. Further, each of these contexts include temporal information about the evolution of interactions. However, none of the measures presented here explicitly handle temporal or weighted hypergraphs, though we comment that Ref.~\cite{barrett_counting_2025} touches on this topic.
The creation of a suite of measures or a procedure for dealing with these features based on existing measures could be translated across different domains, particularly neuroscience, where recent work has inferred higher-order networks using information theory~\cite{varley_partial_2023} and nonlinear system identification~\cite{delabays_hypergraph_2025}, and ecology, such as plant-pollinator networks~\cite{young_reconstruction_2021}.
This perspective would also complement recent developments on weighted simplicial complexes~\cite{baccini_weighted_2022}, for example.

Finally, the encapsulation DAG opens the door to more sophisticated measures of nestedness. While recent work has begun to correlate nestedness with dynamical processes~\cite{kim_contagion_2023,burgio_triadic_2024}, there remains space for dynamics-informed measures, analogous to the hypergraph assortativity introduced in Ref.~\cite{landry_hypergraph_2022}.

\section*{Data and code availability}

All code necessary to reproduce all plots is on \href{https://github.com/tlarock/hypergraph-nestedness}{GitHub}. All datasets are openly available on the XGI-DATA repository~\cite{landry_xgi_2023}.

\begin{acknowledgement}
R.L. acknowledges support from the EPSRC grants EP/V013068/1, EP/V03474X/1 and EP/Y028872/1. N.W.L. acknowledges support from the University of Virginia Prominence-to-Preeminence (P2PE) STEM Targeted Initiatives Fund, SIF176A Contagion Science.
\end{acknowledgement}

\printbibliography
\end{document}